\documentstyle[12pt,aaspp4]{article}
\begin{document}
\parindent=1.0cm

\title{THE EVOLVED RED STELLAR CONTENT OF M32}

\author{T. J. Davidge \altaffilmark{1}}

\affil{Canadian Gemini Office, Herzberg Institute of Astrophysics,
\\National Research Council of Canada, 5071 W. Saanich Road,\\Victoria,
B.C. Canada V8X 4M6\\ {\it email:tim.davidge@hia.nrc.ca}}

\altaffiltext{1}{Visiting Astronomer, Canada-France-Hawaii Telescope, which is
operated by the National Research Council of Canada, the Centre National de la 
Recherche Scientifique, and the University of Hawaii}

\begin{abstract}

	Near-infrared images obtained with the Canada-France-Hawaii Telescope 
(CFHT) Adaptive Optics Bonnette (AOB) are used to investigate the stellar 
content of the Local Group compact elliptical galaxy M32. Observations of a 
field 2.3 arcmin from the galaxy center reveal a large population of asymptotic 
giant branch (AGB) stars, and comparisons with models indicate that these 
objects have an age log(t$_{Gyr}) \leq 9.3$. The AGB population is very 
homogeneous, with $\Delta$ log(t$_{Gyr}) \leq \pm 0.1$ dex and $\Delta$[M/H] 
$\leq \pm 0.3$ dex. The reddest AGB stars have $J-K \leq 1.5$, and it is 
suggested that the very red stars seen in earlier, less deep, 
surveys are the result of large photometric errors. 
The bolometric AGB luminosity function (LF) of this field is in excellent 
agreement with that of the Galactic bulge. Based on the integrated brightness 
of AGB stars brighter than the RGB-tip, which occurs at $K = 17.8$, it is 
concluded that intermediate-age stars account for roughly 25\% of the total $K$ 
light and $10\% \pm 5\%$ of the total mass in this field. 

	A field close to the center of M32 was also observed. The brightest 
stars within a few arcsec of the nucleus have $K = 15.5$, and the density of 
these objects is consistent with that predicted from the outer regions 
of the galaxy after scaling according to surface brightness. 
Moreover, the $K$ luminosity function (LF) of bright sources between 20 and 30 
arcsec of the nucleus is well matched by the LF of the outer regions of the 
galaxy after accounting for differences in surface brightness and correcting 
for the effects of crowding. It is concluded that the relative size of the 
intermediate age component with respect to other populations does not change 
with radius over much of the galaxy. However, the integrated $J-K$ color and 
$2.3\mu$m CO index change with radius within a few tenths of an arcsec of the 
galaxy center indicating that, contrary to what might be inferred from 
observations at visible wavelengths, the integrated photometric properties 
of the central regions of M32 differ from those of the surrounding galaxy.

\end{abstract}

\keywords{galaxies: individual (M32) -- galaxies: Local Group -- 
galaxies: stellar content -- stars: AGB and post-AGB}

\section{INTRODUCTION}

	The Local Group compact elliptical galaxy M32 has long been recognized 
as an important stepping-stone for interpreting the stellar content of more 
distant galaxies. The structural characteristics of M32 appear to be related to 
massive classical ellipticals (e.g. Kormendy 1985). Nevertheless, M32 
is also the closest example of a class of compact elliptical galaxies 
that are junior partners in hierarchal systems (Prugniel, Nieto, \& Simien 
1987), suggesting that the high central concentration may be the result of 
tidal interactions (e.g. Faber 1973). On the other hand, Nieto \& Prugniel 
(1987) find that the separation between M31 and M32 decays at a rate of $2 - 
3$ kpc Gyr$^{-1}$. As a result, the interactions between M31 and M32 were 
likely less intense in the past, and Nieto \& Prugniel (1987) conclude that the 
structural characteristics and mass of M32 have not changed dramatically with 
time. While it is a matter of debate whether or not M32 and 
classical ellipticals share a common evolutionary heritage, it is 
clear that there is little hope of understanding the 
stellar contents of more distant galaxies if we can not understand the 
stellar content of a nearby, resolved system like M32.

	M32 has been the target of numerous spectroscopic investigations, and 
these have detected signatures of a relatively blue main sequence turn-off 
(O'Connell 1980, Rose 1985), and an extended asymptotic giant branch (AGB) 
(Davidge 1990), both of which are indicative of an intermediate-age population. 
M32 also has a very red ultraviolet/visible color when compared with 
other elliptical galaxies (Burstein et al. 1988), and this can be explained 
if a large intermediate age population is present (e.g. Ohl et al. 
1998, Yi et al. 1999).

	During the past decade there has been increased emphasis on studies of 
the resolved stellar content of M32. Imaging of the outer regions of M32 at 
visible wavelengths failed to detect an extended AGB (Freedman 1989, Davidge \& 
Jones 1992; Grillmair et al 1996). This is not unexpected, as the visible 
light from highly evolved metal-rich giants is affected by line blanketing 
(Bica, Barbuy, \& Ortolani 1991), with the result that metal-rich AGB-tip 
stars can be significantly fainter at visible wavelengths than less evolved 
metal-poor stars. The effects of line blanketing are greatly reduced at 
wavelengths longward of $1\mu$m (e.g. Davidge \& Simons 1994), and infrared 
imaging surveys have discovered AGB stars as bright as M$_{K} = -9$ in M32 
(Freedman 1992, Elston \& Silva 1992), which are much more luminous than AGB 
stars in Galactic globular clusters. 

	The outer regions of M32 contain stars spanning a range of 
metallicities (Freedman 1989; Davidge \& Jones 1992), with the peak of the 
metallicity distribution function (MDF) occuring between [Fe/H] = -- 0.2 and -- 
0.3 (Grillmair et al. 1996). The MDF of M32 at intermediate radii can not be 
fit with a simple one-zone enrichment model (Grillmair et al. 1996), suggesting 
a complex chemical enrichment history, possibly like that proposed for 
the massive elliptical galaxy NGC 5128 by Harris, Harris, \& Poole (1999).
Long slit spectra (Cohen 1979, Davidge, de Robertis, \& Yee 1990, Davidge 1991, 
Hardy et al. 1994) and the spatial distribution of resolved and unresolved UV 
sources (Brown et al. 1998) indicate that population gradients are present, 
with mean metallicity being one of the parameters that likely changes 
with radius. While the central regions of M32 contain strong-lined stars that 
originate from a very metal-rich component (Rose 1994), only a 
modest population of UV-bright sources, which are thought to have evolved 
from the extreme horizontal branch (EHB) of a very metal-rich population, 
have been detected in M32, and Brown et al. (1998) conclude that the 
progenitors of these objects account for only 0.5\% of all main sequence stars. 

	In the present study, high angular resolution near-infrared images of 
two fields, sampling different regions of the galaxy, are used to investigate 
the evolved red stellar content of M32. In addition to probing highly evolved 
stars, observations at infrared wavelengths also provide important information 
for decoupling the effects of age and metallicity in 
integrated light measurements (Davidge 1991). Finally, the contrast 
between the reddest, most evolved, stars and the unresolved, relatively blue, 
body of the galaxy is enhanced in the infrared with respect to the visible, 
making it easier to resolve evolved stars in crowded fields. 

	The observations, the data reduction procedures, and the photometric 
measurements are discussed in \S 2. The stellar contents of the two fields are 
investigated in \S 3, and it is demonstrated that the density of the brightest 
AGB stars scales with surface brightness at distances in excess of a few arcsec 
from the nucleus, indicating that the relative size of the intermediate age 
component measured with respect to the rest of the galaxy is constant 
over a large range of radii. However, within 1 arcsec of the nucleus 
the integrated near-infrared broad-band colors and the $2.3\mu$m CO index vary 
with radius. A brief summary and discussion of the results follows in \S 4.

\section{OBSERVATIONS, REDUCTIONS, AND PHOTOMETRIC MEASUREMENTS}

	The data were recorded with the Canada-France-Hawaii Telescope (CFHT) 
Adaptive Optics Bonnette (AOB) and KIR imager on the nights 
of UT September 6 and 7, 1998. The AOB has been 
described by Rigaut et al. (1998). The detector in KIR is a $1024 \times 1024$ 
Hg:Cd:Te array with 0.034 arcsec pixels, so that the imaged field is $34 
\times 34$ arcsec. 

	A field that included the nucleus of M32 (the `inner' field) was 
observed through $J, H, Ks, 2.2\mu$m continuum, and $2.3\mu$m CO filters. The 
galaxy nucleus, which was used as the reference source for AO compensation, 
was offset 10 arcsec from the field center so that the minor axis of 
M32 could be sampled out to a distance of 40 arcsec. 
A complete observing sequence for each filter consisted of 5 co-added 
exposures recorded at 4 dither positions, which were offset in 
a $0.5 \times 0.5$ arcsec square pattern. The total exposure times were 
1200 sec in each of $J, H,$ and $Ks$, 1800 sec in the $2.2\mu$m continuum 
filter, and 2400 sec in CO. The seeing conditions were mediochre when these 
data were recorded, and the delivered image quality is 0.4 arcsec FWHM.

	Deep $J, H,$ and $Ks$ images, with 1200 sec total exposure times per 
filter, were also recorded of a field 2.3 arcmin from the nucleus (the 
`outer' field) centered on the star GSC 02801 -- 02082, which was used as 
the reference source for AO compensation. This field is slightly offset from 
the minor axis of the galaxy, and the data were recorded using the observing 
sequence described for the inner field. 
The seeing conditions were very good, and the delivered 
image quality is 0.20 arcsec FWHM in $J$ and 0.15 arcsec in $H$ and $Ks$.

	The data were reduced using the procedure described by Davidge \& 
Courteau (1999), which consists of the following steps: (1) subtraction 
of dark frames, (2) division by dome flats, (3) subtraction of a DC sky level 
from each image, and (4) subtraction of fringe and thermal emission patterns, 
using calibration frames that were constructed from blank sky fields. The 
images for each field were then aligned and median-combined on a 
filter-by-filter basis. The final $Ks$ images of these fields are shown in 
Figures 1 and 2.

	Stellar brightnesses were measured with the PSF-fitting program ALLSTAR 
(Stetson \& Harris 1988), using PSFs and target lists generated with DAOPHOT 
(Stetson 1987). A single PSF was constructed for each field $+$ filter 
combination. The use of a single reference star for AO-compensation produces a 
variable PSF (`anisoplanaticism'), and this is a source of photometric errors 
if a single PSF is adopted for the entire field. However, studies 
of globular clusters conducted with the CFHT AOB demonstrate that the 
photometric uncertainties introduced by PSF variations 
amount to only a few percent over the KIR field 
(Davidge \& Courteau 1999; Davidge 2000a), and an examination of 
the PSF-fitting residuals in the inner and outer field M32 data indicate that 
PSF variations do not introduce uncertainties of more than a few percent. The 
modest scatter in the M32 outer field CMDs (\S 3) is further evidence that PSF 
variations do not seriously affect the current photometry. 

	The photometric calibration was defined using standard 
stars from Casali \& Hawarden (1992) and Elias et al. (1982), which were 
observed throughout the 3 night run. The unresolved body of M32 creates a 
non-uniform background in the inner field that complicates efforts to measure 
stellar brightness, and this was removed using the iterative procedure 
described by Davidge, Le Fevre, \& Clark (1991).

\section{RESULTS}

\subsection{The Outer Field}

	The $(K, H-K)$ and $(K, J-K)$ CMDs of the M32 outer field 
are shown in Figure 3. The RGB and upper portions of the AGB are clearly 
evident in these data: the onset of the former causes a slight broadening in 
the color distribution when $K \geq 17.8$, while the latter extends from $K = 
17.5$ to $K = 16$, and has a scatter of $\pm 0.05$ mag in $H-K$ and $J-K$. 
There is also a 0.2 mag gap in the CMDs near $K = 17.6$.

	Data from Table 1 of Freedman (1992) are compared with the current 
observations in the right hand panel of Figure 3. Freedman observed an area 
four times larger than the AOB field, and so it is not surprising that she 
detected sources that are brighter than those in the present dataset. 
The CMDs constructed from the AOB data have less scatter than the Freedman 
observations, and extend $1 - 1.5$ mag fainter in $K$. The ridgeline of the 
Freedman observations, which can be traced by identifying the colors 
with the highest concentration of data points, is located near 
the left hand side of her data distribution and matches the locus of the AOB 
data. A gap in the CMD immediately above the RGB-tip is also evident in the 
Freedman data.

	Freedman suggested that the objects in her dataset with $J-K \geq 1.5$ 
are carbon stars. Stars with this color are not seen in the current dataset, 
even though a significant number would have been expected based on the Freedman 
observations. While stars with $J-K \geq 1.5$ are present in the Elston \& 
Silva (1992) $(K, J-K)$ CMD, the relative number of these objects with 
respect to bluer stars is much smaller than in the Freedman (1992) dataset. 
Freedman (1992) notes that some stars in her study have large photometric 
uncertainties, which may be as great as 0.5 mag in $J-H$ and 0.2 mag 
in $H-K$. We speculate that the reddest stars 
in her sample are objects with large photometric errors. 

	The $K$ LF of the outer field is plotted in Figure 4. Artificial star 
experiments indicate that incompleteness and photometric errors become 
significant when $K = 20$, and so only data brighter than this are 
shown in Figure 4. The dashed line shows a power-law that was fit to the LF 
between $K = 18$ and $K = 19.5$ using the method of least squares. The fitted 
power-law has an exponent $0.16 \pm 0.06$, which is near the low end of what is 
seen in the Galactic bulge (e.g. Davidge 2000b). When $K \leq 18$ the LF 
consistently falls below the extrapolated power-law fit, 
indicating that the RGB-tip produces a significant discontinuity in the $K$ LF, 
even though models predict that the brightness of the RGB-tip is sensitive to 
age and metallicity near $2\mu$m (see below).

	The bolometric LF of stars in the outer field, calculated using the 
bolometric corrections for field giants shown in Figure 1b of Frogel \& 
Whitford (1987), is plotted in Figure 5. A distance modulus of $\mu_0 = 24.3$ 
has been adopted, based on an RGB-tip brightness of $I = 20.5$ for M32 (Davidge 
\& Jones 1992), which is 0.1 mag brighter than that in M31 (Davidge 1993), for 
which $\mu_0 = 24.4 \pm 0.1$ (van den Bergh 2000). The data have also been 
corrected for a foreground reddening of E(B--V) = 0.08 (Burstein \& Heiles 
1984).

	The most luminous stars in the M32 outer field have M$_{bol} \sim -6$, 
which is comparable to what is seen in Baade's Window (e.g. Frogel \& Whitford 
1987). The dashed line in Figure 5 is the LF of inner bulge giants measured by 
Davidge (1998), which has been scaled to match the number of stars in the M32 
field with M$_{bol} \leq -3.5$. There is excellent agreement between the bulge 
and M32 LFs. 

	The brightnesses of the AGB-tip and RGB-tip from the z = 0.001 
([M/H] = --1.2), z =  0.004 ([M/H] = --0.6), and z = 0.020 ([M/H] = 0.1) 
Bertelli et al. (1994) models are compared with the observations in 
Figure 6. M$_K$ was computed from the M$_V$ and 
$V-K$ entries in Tables 8, 9, and 11 of Bertelli et al. (1994), while 
$J-K$ colors were calculated using the field giant 
color relations listed in Table 3 of Bessell \& Brett (1988). 

	The model sequences in Figure 6 consistently fall to the right of the 
outer field observations, although the significance of this result is low 
due to uncertainties in, for example, the mixing length and the adopted 
relation between $V-K$ and $J-K$. If the distance to M32 has been overestimated 
then the agreement along the color axis will improve slightly. While there are 
uncertainties in the color calibration, the relative model-to-model color 
variations are likely reliable, and the scatter on the upper AGB is consistent 
with a dispersion of $\pm 0.3$ dex in [M/H] or $\pm 0.1$ dex in age.

	Comparisons between models and observations that are based on 
brightness are more reliable than those that rely on color. The inner field 
observations suggest that the AGB-tip occurs near $K = 15.5$ in M32 (\S 3.2), 
which is consistent with the brightest stars detected in the 
Freedman (1992) and Elston \& Silva (1992) datasets, and corresponds to 
M$_{K} = -8.8 \pm 0.1$ if $\mu_0 = 24.3 \pm 0.1$. 
It is evident from Figure 6 that models with log(t$_{Gyr}$) 
$\leq 9.3$ are required to match the AGB-tip brightness. 
Caution should be exercised when attempting to derive absolute ages 
from a highly evolved state such as the AGB-tip, and 
the age limits inferred from Figure 6 are uncomfortably 
close to those defined by the visible spectrum of M32, which indicate that the 
galaxy does not contain a large population with log(t$_{Gyr}$) $\leq 9.0$ 
(Bica et al. 1990).

\subsection{The Circumnuclear Region}

	In this section the stellar content in the inner M32 field 
is compared with that in the outer field. Mean age and/or metallicity 
may change rapidly with radius near the center of M32, and so the 
inner field was divided into three regions, 
sampling distinct stellar density regimes centered on the nucleus. 
The radial limits and the mean surface brightnesses of these regions 
are listed in Table 1. The effects of crowding in 
Region 1 are extreme, and the stellar content in this area will be 
investigated using integrated colors in \S 3.3. 

	The $K$ LFs of Regions 2 and 3 are shown in Figure 7. The lower stellar 
density in Region 3 means that the brightness in $K$ where incompleteness 
becomes significant, estimated from the point at which number counts in the 
LFs decline, is 0.5 -- 1.0 mag fainter in Region 3 than in Region 2. 

	Many of the sources detected in Regions 2 and 3 are not 
individual stars, but blends of two or more moderately faint objects. 
The effects of blending become less severe with increasing brightness, 
and the densities of stars with $K = 16$ are sufficiently low that 
the sources with $K = 15.5$ in Regions 2 and 3 are likely 
individual stars. This can be checked by comparing the densities of $K = 15.5$ 
stars in the Freedman (1992) and Elston \& Silva (1992) datasets with those 
in Region 2 and 3 after correcting for differences in surface brightness.

	The observed densities of $K = 15.5$ stars in the inner field are $240 
\pm 50$ per square arcmin (Region 2) and $14 \pm 8$ per square arcmin (Region 
3). Freedman (1992) detected 4 stars with $K$ between 15.75 and 15.25 in a 1.11 
square arcmin area 2 arcmin south of the M32 nucleus, where the surface 
brightness is $\mu_r = 21.9$ mag per square arcmin (Kent 1987). If the spatial 
distribution of $K = 15.5$ stars follows the integrated light profile of the 
galaxy then the densities of these objects in the circumnuclear region should 
be $200 \pm 100$ per square arcmin (Region 2) and $50 \pm 25$ per square arcmin 
(Region 3). Elston \& Silva (1992) detected 10 stars with $K$ between 15.25 
and 15.75 in a 16 square arcmin field 3 arcmin east of the nucleus, where the 
surface brightness is $\mu_r = 23.1$ mag per square arcmin. The densities of 
$K = 15.5$ stars predicted from these data is $110 \pm 11$ per square arcmin 
(Region 2) and $27 \pm 3$ per square arcmin (Region 3). The bright star 
densities predicted for Regions 2 and 3 from the Freedman and 
Elston \& Silva studies agree within the estimated uncertainties, and 
the unweighted means are $150 \pm 50$ (Region 2) and 
$39 \pm 13$ (Regions 3). Thus, the observed densities of 
bright red stars in Regions 2 and 3 differ from the densities inferred from 
the outer regions of M32 at less than the $2\sigma$ level. 

	The calculations described above indicate that the 
the brightest red stars have a spatial distribution that follows the 
surface brightness profile of the galaxy to within a few arcsec 
of the nucleus. These comparisons can be extended to fainter stars by 
scaling the outer field LF to match the surface brightnesses 
in Regions 2 and 3. The effects of crowding are significant 
among all but the brightest sources in Regions 2 and 3, and statistical 
crowding corrections appropriate for each Region were applied to the outer 
field LF after calculating the probability that two stars in a given 
brightness interval occur in the same resolution element, the diameter 
of which equals the FWHM of the PSF, and thus 
produce a blended image that would be counted in the next brightest LF bin.

	The outer field LF, scaled and corrected for crowding based on the 
stellar densities in Regions 2 and 3, is shown as a dashed line in Figure 7. 
While the agreement with the Region 3 LF is excellent when $K \leq 17$, 
the comparison with the Region 2 LF is inconclusive, as incompleteness 
becomes significant in Region 2 when $K > 16$. The 
data thus indicate that at distances in excess of 20 arcsec from 
the center of M32, which is the inner radius of Region 3, the 
evolved red stellar content scales with $\mu_r$. Therefore, 
the intermediate-age population is not more centrally 
concentrated than other populations at distances in excess of 20 arcsec 
from the nucleus. While the density of $K = 
15.5$ stars suggests that this may also be the case for radii as small as 
5 arcsec, which is the inner radius of Region 2, this can not be verified 
with fainter stars until data with better image quality are obtained.

	Although M32 contains a metallicity gradient (e.g. Hardy et al. 1994), 
this will not have a major impact on the comparisons discussed 
above, as the radial changes in mean metallicity are modest. 
If $\Delta$Mg$_2$/$\Delta$log(r) = $-0.030 \pm 0.005$ and 
$\Delta<Fe>/\Delta$log(r) = $-0.57 \pm 0.11$ (Hardy et al. 1994), then 
the difference in indices between the outer field and Region 3 
($\Delta$log(r) = --0.7) is $\Delta$Mg$_2 = 0.02$ and $\Delta<Fe> = 0.36$, 
while between the outer field and Region 2 ($\Delta$log(r) = --1.0) 
$\Delta$Mg$_2 = 0.03$  and $\Delta<Fe> = 0.57$. These correspond 
to $\Delta$[Fe/H] = 0.1 dex (Region 3) and $\Delta$[Fe/H] = 0.2 dex (Region 2) 
with respect to the outer field according to the solar metallicity models of 
Worthey (1994). 

\subsection{The Nucleus of M32}

	The radial behaviour of $J-H$, $J-K$ and CO near the center of M32 is 
investigated in Figure 8, where radial averages of these quantities in 0.4 
arcsec annuli are plotted. $J-K$ becomes bluer when $r \leq 1$ arcsec, while 
the CO index becomes larger. Given that $J-H$ does not change markedly with 
radius, it appears that the color gradients are driven primarily by changes in 
the spectral-energy distribution (SED) at wavelengths longward of $\geq 2\mu$m. 
The colors in Figure 8 are insensitive to uncertainties in sky 
brightness, which was measured in the upper left hand corner of Figure 2,
because they are restricted to the high surface brightness central regions of 
the galaxy. 

	Peletier (1993) used images with an angular resolution of 
1.5 arcsec FWHM to investigate near-infrared 
color gradients in M32. The data in Table 2 of Peletier (1993) indicate that 
within the central few arcsec of M32 there is a clear tendency for $J-K$ 
to weaken with decreasing radius, and for CO to strengthen, in 
qualtitative agreement with the trends in Figure 8.

	Several studies have measured the visible and UV 
colors near the center of M32. Michard \& Nieto (1991) examined data with 
0.7 -- 0.9 arcsec resolution and concluded that $B-R$ and $U-B$ do not vary 
with radius near the nucleus of M32. However, the measurements listed in 
Table 2 of Peletier (1993), indicate that $U-R$ and $B-R$ 
may redden towards smaller radii. Lauer et al. (1998) examined images recorded 
with WFPC2 and concluded that $V-I$ does not change with radius, while $U-V$ 
may redden with decreasing radius. Ohl et al. (1998) find that the color of M32 
in the far-ultraviolet becomes significantly redder towards smaller radii, 
although this trend is defined with measurements at distances in excess of 
2 arcsec from the center, and so may not be related to changes in stellar 
content near the nucleus. The published data indicate that visible and 
near-infrared colors do not change dramatically with radius in M32 when $r \geq 
1$ arcsec, and none of the studies detect a central bluing in the UV and 
visible portions of the spectrum. When combined with the results in Figure 8, 
the published data suggest that the SED of M32 undergoes complex changes 
close to the nucleus, becoming bluer with decreasing radius longward 
of 2$\mu$m, not changing with radius at visible wavelengths, and possibly 
becoming redder with decreasing radius shortward of $0.4\mu$m. 

	The $(J-H, H-K)$ and $(CO, J-K)$ diagrams, shown in Figure 9, provide a 
means of probing the SED near the nucleus of M32. When $r \geq 1$ arcsec the 
near-infrared SED is similar to that of star clusters in the Magellanic Clouds. 
However, at smaller radii the M32 data depart from a SED appropriate for star 
clusters, in the sense that while $J-H$ remains at a value appropriate for 
stellar systems, $H-K$ decreases by more than 0.1 mag. When $r \leq 1$ arcsec 
$J-K$ decreases towards smaller radii, while the CO index increases.

	The large nuclear CO index of M32 is not easily explained. 
Mobasher \& James (1996), and James \& Mobasher (1999) measured the 
strength of $2.3\mu$m CO absorption in a moderately large number of elliptical 
galaxies. These data were recorded through 2.4 arcsec wide slits, and hence are 
not dominated by light from the galaxy nuclei; nevertheless, these data still 
provide insight into the CO index of composite metal-rich systems. 
After transforming the spectroscopic CO measurements into the photometric CO 
system using the relation defined by Doyon, Joseph, \& Wright (1994), it is 
concluded that CO$_{max} = 0.21$, compared with CO = 0.36 near the nucleus of 
M32. The only other stellar system having a published CO index with a 
spatial resolution comparable to the M32 observations is the Galactic 
Center, which would have CO = 0.26 if viewed 
at a distance of 1 Mpc (Davidge 2000c).

	Luminous red giants in the vicinity of SgrA and red supergiants in the 
Galaxy and the Magellanic Clouds have CO indices comparable to 
the nucleus of M32 (Elias et al. 1985, Kenyon 1988, Davidge 1998). 
However, these stars have $J-K > 1$, and hence do not provide 
suitable templates for the nucleus of M32. Therefore, it appears that the 
near-infrared SED of the M32 nucleus differs from that of benchmark systems 
such as nearby bright giants, star clusters, and the central few arcsec of 
elliptical galaxies.

\section{SUMMARY \& DISCUSSION}

	Near-infrared images have been used to investigate the stellar content 
of the Local Group compact elliptical galaxy M32. 
AGB stars as bright as $K = 15.5$, which 
were detected in earlier infrared studies of the outer regions of the galaxy, 
are also found near the center of the galaxy. The number density of these 
objects increases towards smaller radii, indicating that they are actual 
members of M32, and not luminous AGB stars in the outer disk of M31. 
Hence, these data confirm the presence of the intermediate-age 
population predicted by spectroscopic investigations. The bright 
AGB stars in the outer regions of M32 have $J-K \leq 1.5$, and it is suggested 
that the very red objects detected by Freedman (1992) have 
large photometric errors.

	The $K$ LF of the brightest AGB stars scales with integrated surface 
brightness to within 20 arcsec of the galaxy center, suggesting that 
the relative frequency of bright AGB stars, as measured with respect to 
fainter stars, does not change over most of the galaxy.
This does not rule out the presence of a subtle ($\pm 1$ Gyr) age gradient 
in M32, since the brightness of the AGB-tip changes slowly with age in 
all but the youngest intermediate-age populations. The UV portion of the 
spectrum is very age-sensitive during intermediate epochs (Yi et al. 
1999), and an age gradient, in the sense that younger stars occur at smaller 
radii, provides one explanation for the UV color of M32 increasing with 
decreasing radius (O'Connell et al. 1992).

	The data presented in this paper indicate that the near-infrared SED 
changes within a few tenths of an arcsec of the center of M32, and that 
the center of the galaxy has colors that differ from those of star clusters 
and the main bodies of elliptical galaxies. The central properties of M32 
are peculiar in other ways. For example, the ratio of central x-ray flux to 
mass is the smallest of any known super-massive black hole, suggesting that the 
central object may be fuel-starved, or is not an efficient accretor 
(Loewenstein et al. 1998). In addition, the gravitational field from the 
central object influences the central arcsec of M32 (van der Marel 1999), so 
variations in stellar content and/or morphology might be expected to occur in 
this area. However, the core of M32 lacks conspicuous structure at UV 
(Cole et al. 1998) and visible (Lauer et al. 1998) wavelengths. M32 also lacks 
a centrally concentrated blue population, and this is puzzling 
since stellar interactions in the dense nuclear regions might be expected to 
produce a large number of blue stragglers (Lauer et al. 1998). The changes in 
near-infrared photometric properties detected in the current study occur on an 
angular scale that is comparable to the region influenced by the 
central black hole (van der Marel 1999), suggesting a possible connection. 
The `stubborn refusal of M32 to reveal any departures from utter normalacy at 
small radii' noted by Lauer et al. (1998) appears to break down when data 
spanning a broader range of wavelengths are considered.

	The relative size of the intermediate-age population with respect to 
the rest of the galaxy can be estimated from the 
integrated brightness of AGB stars brighter than the RGB-tip. The integrated 
brightness of the outer field based on the Kent (1987) surface brightness 
profile, assuming that $V - r = 0.2$ for a typical spheroidal galaxy 
(Kent 1987) and that $V - K = 3.1$ for M32 (Frogel et al. 1978), is $K = 11.8$. 
For comparison, the integrated brightness of stars on the upper AGB is $K = 
13.5$, so luminous AGB stars contribute 20\% of the total light from the 
outer field in $K$.

	Frogel, Mould, \& Blanco (1990) studied the near-infrared photometric 
properties of AGB stars in a sample of SMC and LMC clusters, and 11 of these 
clusters (NGCs 411, 419, 1751, 1806, 1846, 1978, 2108, 2121, 2154, 2213, and 
2231) have SWB types V, V-VI, and VI, indicating that they have ages of a few 
Gyr, and integrated $K$ brightness measurements 
(Persson et al. 1983). The ratio of light in $K$ from bright AGB stars to 
that of the entire cluster is a well-defined quantity among these objects, with 
a mean value $0.81 \pm 0.02$. Therefore, if the intermediate age component in 
the M32 outer field has an age of a few Gyr then this population accounts for 
roughly 25\% of the total light in $K$. The M/L ratio in $K$ of an intermediate 
age population that has a Salpeter IMF and an age of a few Gyr is one-half to 
one-third that of an old population (Buzzoni 1989). If the main body 
of M32 is relatively old, then the intermediate-age population in the outer 
field thus accounts for roughly 10\% of the total mass, with an estimated 
uncertainty $\pm 5\%$. For comparison, Bica et al. (1990) predict that the 
central regions of M32 contain an intermediate-age population that accounts for 
15\% of the mass. The good agreement between these two mass estimates, obtained 
using very different techniques in different parts of the galaxy, is 
further evidence that the intermediate-age population in M32 is uniformly 
distributed throughout the galaxy, as expected based on the comparisons 
in \S 3.2.

	The presence of an intermediate age population accounting for 10\% of 
the galaxy mass indicates that M32 had a significant ISM a few Gyr in the past. 
What was the source of the gas from which these 
stars formed? M31 is a nearby reservoir of star-forming 
material, and there are signs that M32 has interacted recently (Cepa \& Beckman 
1988) with material in the disk of this galaxy (Byrd 1976, 1978; Sofue \& Kato 
1981). However, if M32 was able to cull large amounts of gas from 
M31 during intermediate epochs then it must have been a 
special event, as subsequent passages through the disk of M31 have not 
produced noticeable bursts of star formation.

	Mass ejected from evolved stars is another source of an ISM. Sage, 
Welch, \& Mitchell (1998) estimate that stellar mass loss should have produced 
an ISM in M32 with mass $3 \times 10^{5}$ M$_\odot$ since the last interaction 
with M31. This is consistent with the mass of HI within the central 3 square 
arcmin, which Emerson (1974) measures to be $\leq 6 \times 10^{5}$ M$_\odot$. 
However, based on the non-detection of CO 1 - 0 emission from M32, Sage et al. 
(1998) conclude that the mass of H$_2$ is $\leq 5.1 \times 10^{3}$ M$_\odot$ 
within the central 160 pc of the galaxy. 

	The mean separation between M32 and M31 was significantly greater in 
the past (Nieto \& Prugniel 1987), and M32 could have built up a larger 
ISM from stellar mass loss during earlier epochs due to decreased 
tidal interactions with M31. If the bright AGB stars formed from gas that 
originated in M32 then these objects should be among the most metal-rich in the 
galaxy, with metallicities within a factor of 2 of solar 
according to the Grillmair et al. (1996) MDF. High-quality 
visible and near-infrared spectra can be obtained of the brightest AGB 
stars using instruments on 4 or 8 metre telescopes, thereby providing a 
direct means of measuring the metallicities of these objects. 
If the AGB stars are found to be relatively metal-poor then a 
source of gas external to M32 would seem likely; for example, if star 
formation in M32 during intermediate epochs was fueled by the accretion of 
the high velocity clouds that have been interpreted as the remnants of the 
material from which the Local Group formed, then the AGB stars would have 
[Fe/H] $\leq -1$ (Blitz et al. 1999).

\clearpage

\begin{center}
FIGURE CAPTIONS
\end{center}

\figcaption
[fg1.eps]
{The M32 inner field, as imaged with the CFHT AOB $+$ KIR in $Ks$. North 
is at the top, and East is to the left. The image covers $30 \times 30$ 
arcsec, and the galaxy nucleus is in the lower right hand corner. 
The unresolved background of the galaxy has been subtracted 
using the procedure described by Davidge et al. (1991).}

\figcaption
[fg2.eps]
{The M32 outer field, as imaged with the CFHT AOB $+$ KIR in $Ks$. North 
is at the top, and East is to the left. The image dimensions are roughly 
$30 \times 30$ arcsec. The bright source in the field center is the star 
GSC 02801 -- 02082.}

\figcaption
[fg3.eps]
{The $(K, H-K)$ and $(K, J-K)$ CMDs of the outer field. The RGB-tip 
occurs at $K = 17.8$, and stars brighter than this are evolving on the AGB. 
Data with $K$ and $J-K$ measurements from Table 1 of Freedman (1992) are 
plotted as open squares in the right hand panel.}

\figcaption
[fg4.eps]
{The $K$ LF of the outer field. n$_{05}$ is the number of stars per 0.5 mag 
interval per square arcsec. Incompleteness and photometric errors become 
significant when $K = 20$, and so data fainter than this are not shown. The 
dashed line is a power-law that was fit to the LF entries with $K$ between 18 
and 19.5. The discontinuity near $K = 17.8$ in the lower panel is due to the 
RGB-tip.}

\figcaption
[fg5.eps]
{The bolometric LF of stars in the outer field. n$_{05}$ is the number 
of stars per 0.5 mag interval per square arcsec. Bolometric corrections were 
computed from $J-K$ colors using the field giant calibration of Frogel \& 
Whitford (1987). A distance modulus of 24.3 has been adopted for M32, and the 
data have been corrected for a foreground reddening of E(B--V) = 0.08 (Burstein 
\& Heiles 1984). The dashed line in the lower panel is the LF of 
giants in the inner bulge of the Galaxy derived by Davidge (1998) and scaled to 
match the number of stars with M$_{bol} \leq -3.5$ in the M32 outer field.}

\figcaption
[fg6.eps]
{The $(K, J-K)$ CMD of the outer field, compared with the AGB-tip 
and RGB-tip models for z=0.001, z=0.004, and z=0.020 from Bertelli et al. 
(1994). Models for four ages are shown. The solid lines 
connect the various points for each evolutionary phase, 
while the dashed lines connect the RGB-tip and AGB-tip 
values at each metallicity. The open squares show stars with $K \leq 16$ 
from Table 1 of Freedman (1992).}

\figcaption
[fg7.eps]
{The $K$ LFs of sources in Regions 2 and 3. 
n$_{05}$ is the number of sources per 0.5 mag interval per square 
arcsec, and the errorbars show the 1 $\sigma$ uncertainties predicted from 
counting statistics. The dashed line shows the outer field LF scaled to 
match the stellar density in Regions 2 and 3 using surface brightness 
measurements from Kent (1987). The outer field LFs shown in this figure have 
been adjusted for the effects of crowding expected in Regions 2 and 3 using 
the procedure described in the text.}

\figcaption
[fg8.eps]
{The radial behaviour of $J-H$, $J-K$, and CO near 
the center of M32. The points plotted in this figure are azimuthal averages 
in 0.4 arcsec wide annuli.}

\figcaption
[fg9.eps]
{The $(J-H, H-K)$ and $(CO, J-K)$ diagrams constructed from the data plotted in 
Figure 8. Points at 0.2 and 1.4 arcsec from the nucleus are labelled. 
The open squares are observations of clusters in the Magellanic Clouds from 
Persson et al. (1983), and the M32 data depart from the cluster 
sequence when $r \leq 1.0$ arcsec.} 

\clearpage 

\begin{table*}
\begin{center}
\begin{tabular}{clc}
\tableline\tableline
Region & Radii\tablenotemark{a} & $\overline{\mu_r}$\tablenotemark{b} \\
 & (arcsec) & \\
\tableline
1 & 0 -- 5 & 14.7 \\
2 & 5 -- 20 & 17.5 \\
3 & 20 -- 43 & 19.0 \\
\tableline
\end{tabular}
\end{center}
\caption{Radii and Surface Brightnesses of Regions 1, 2, and 3}
\tablenotetext{a}{Measured from the galaxy center.}
\tablenotetext{b}{Mean surface brightness, from Kent (1987).}
\end{table*}

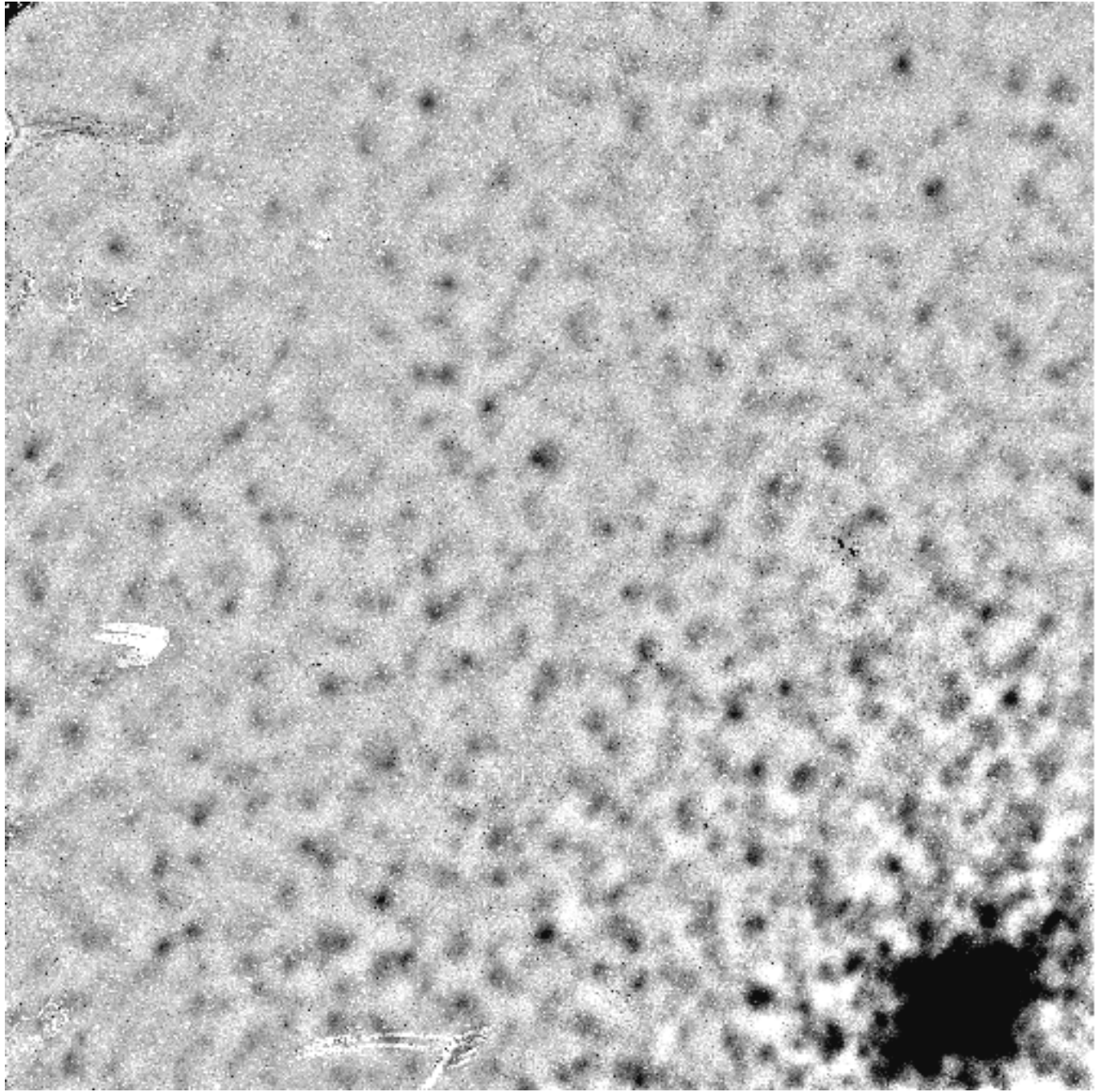
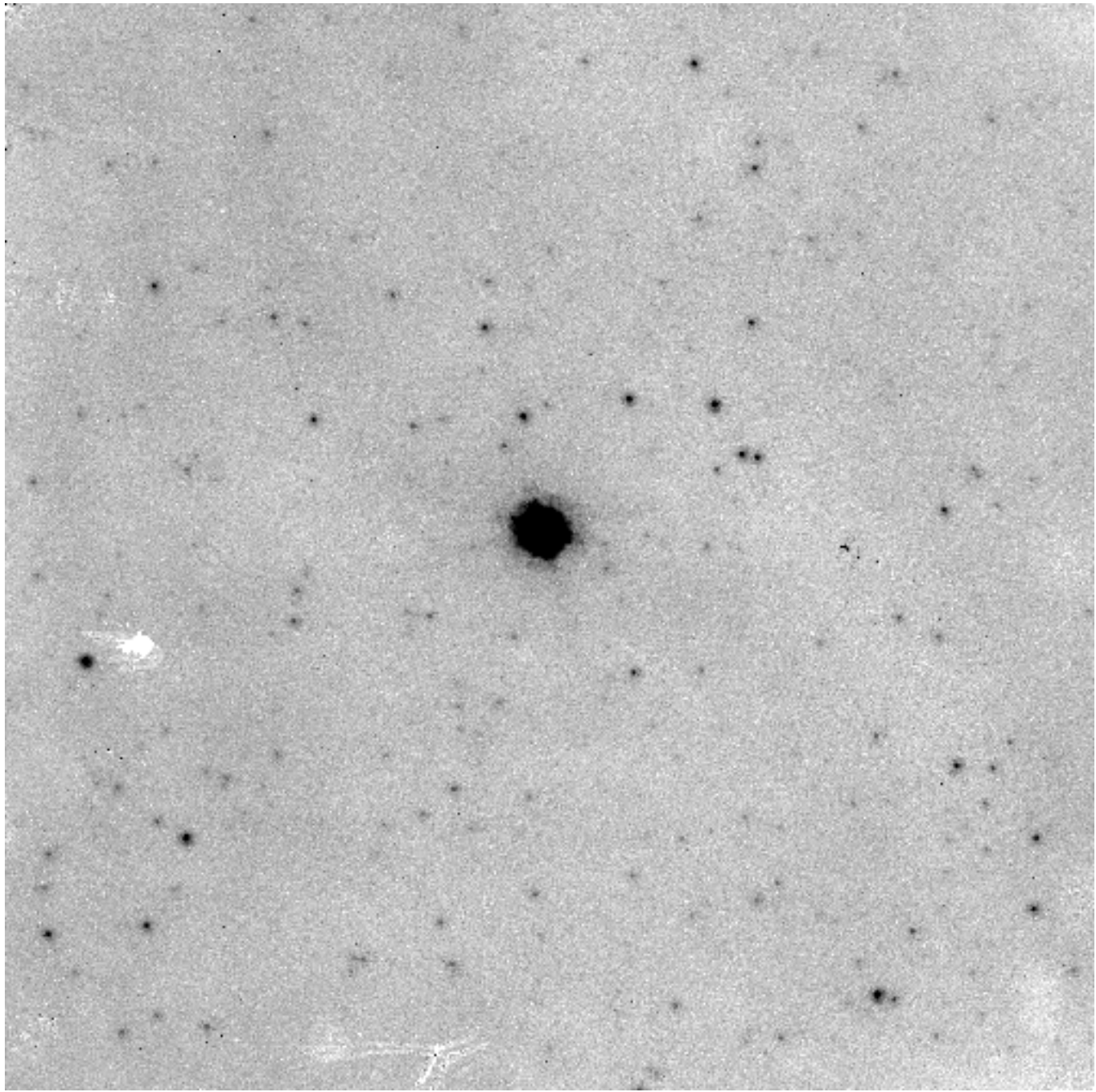
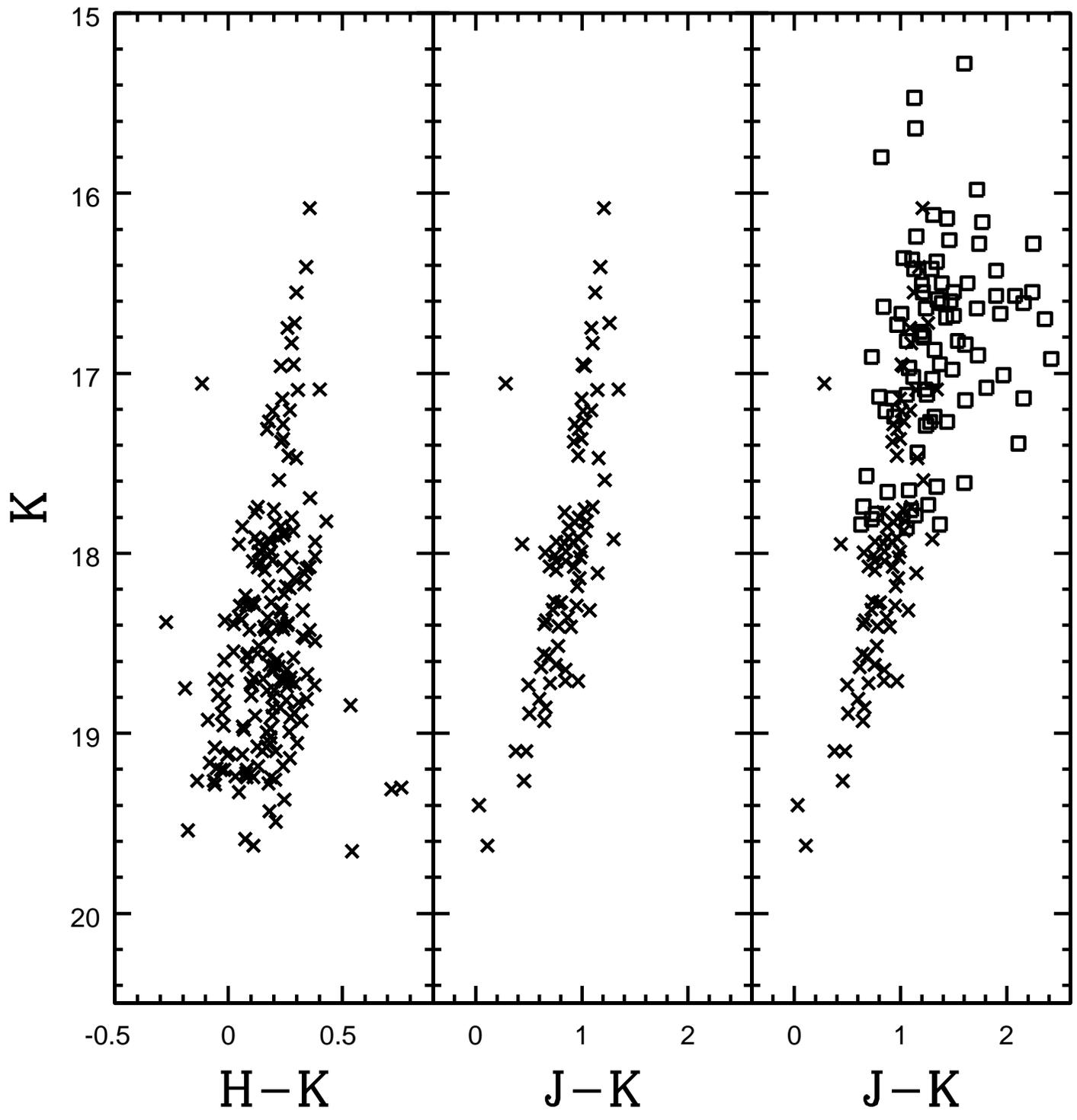
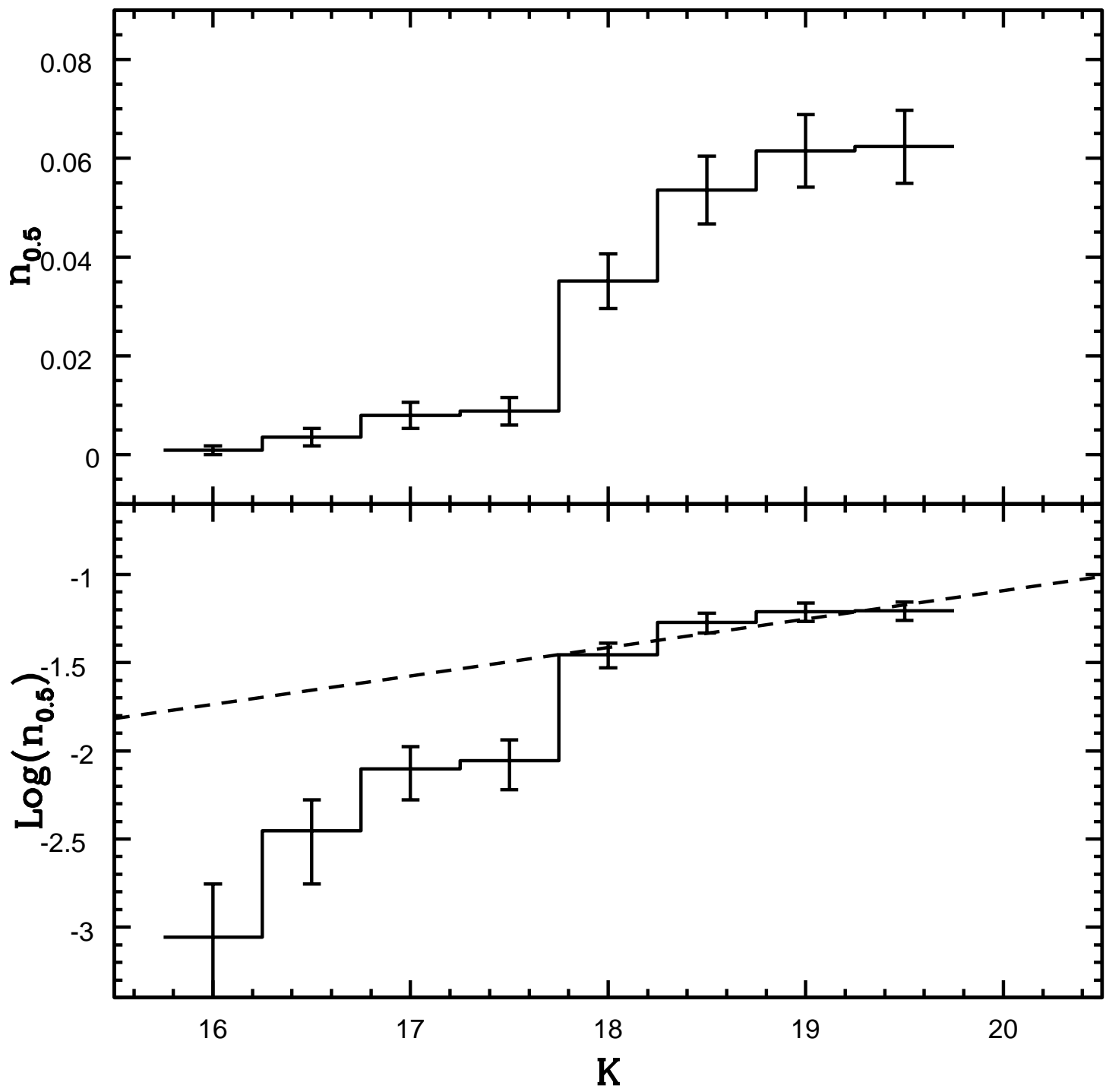
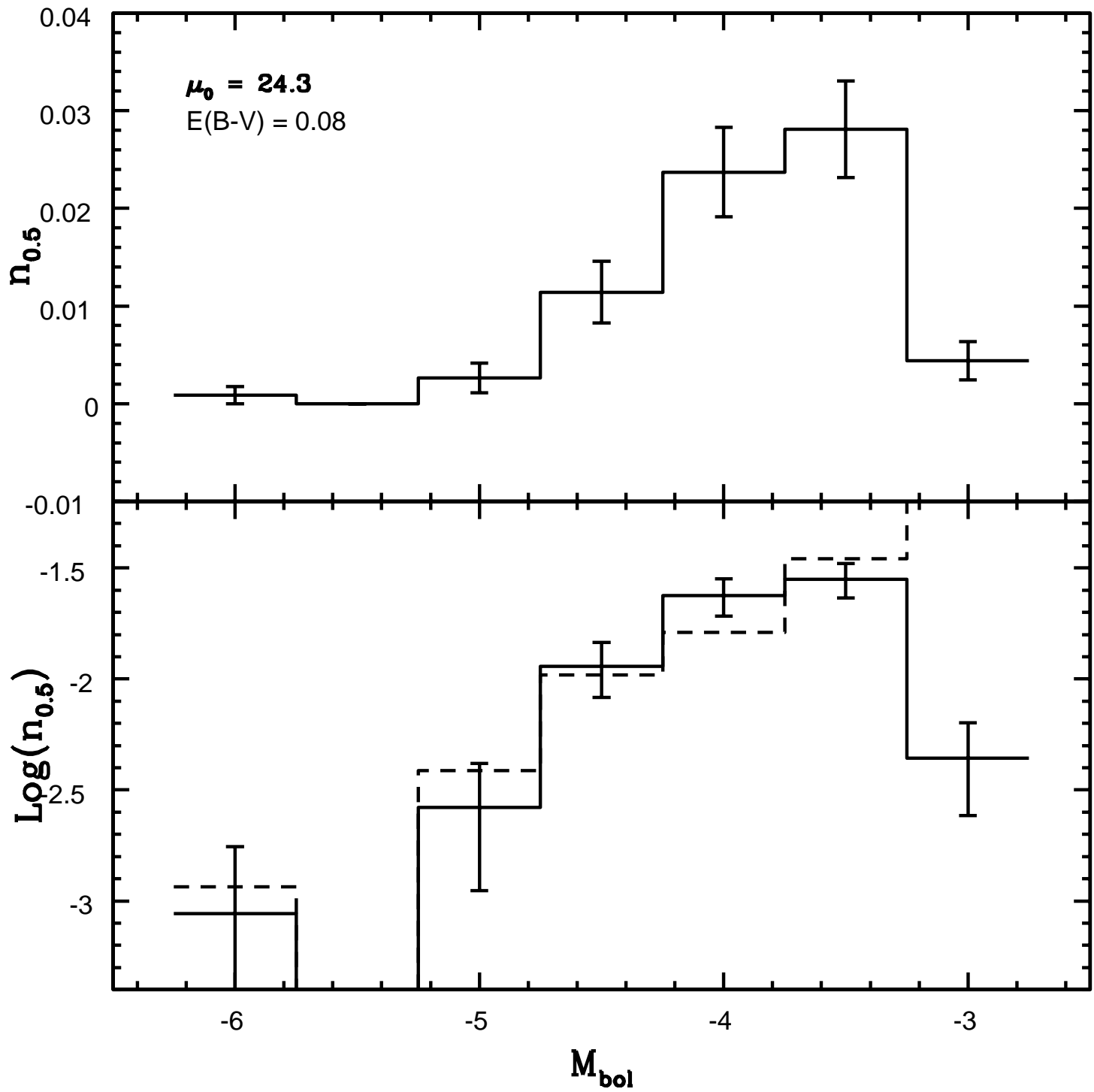
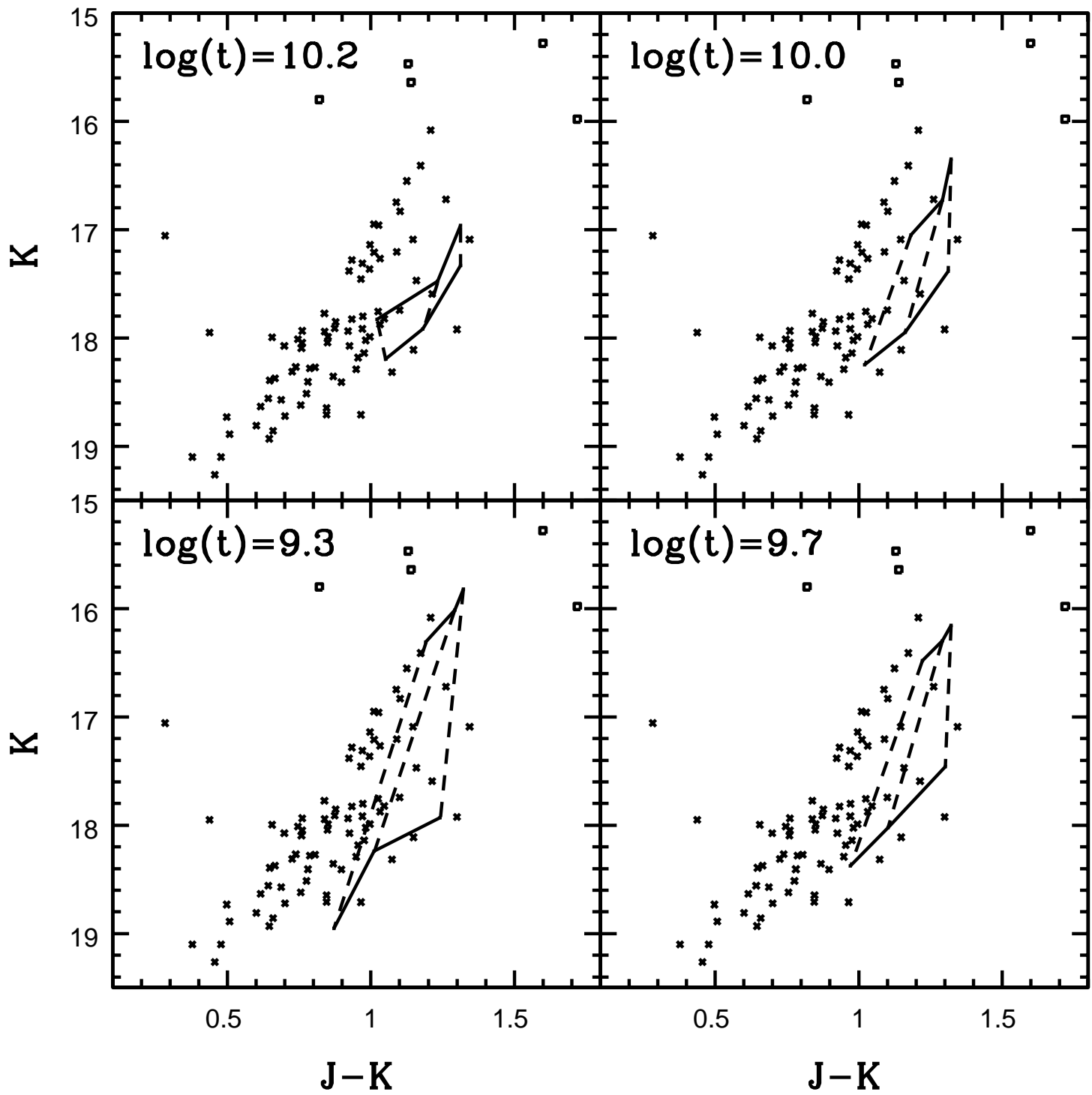
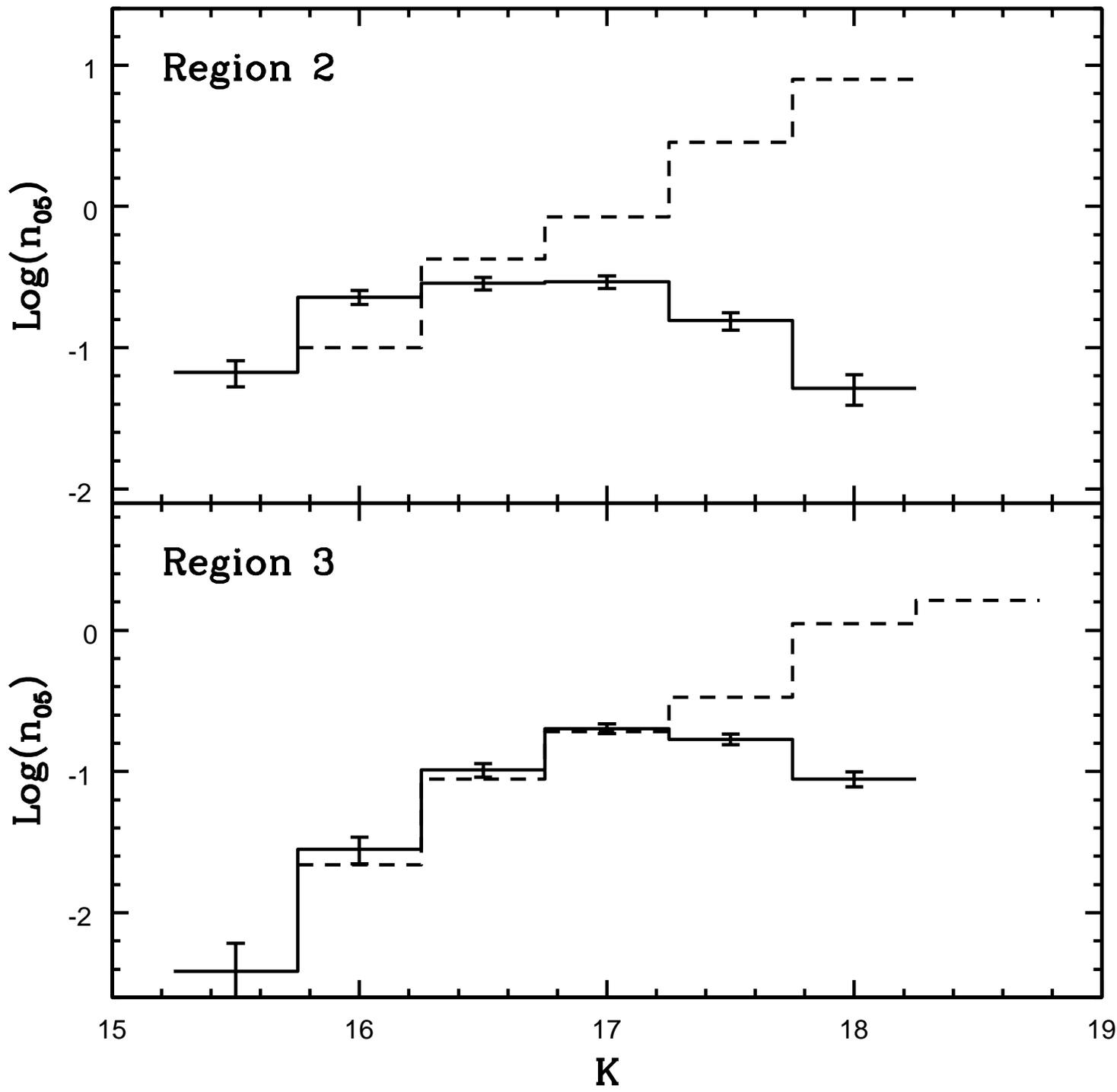
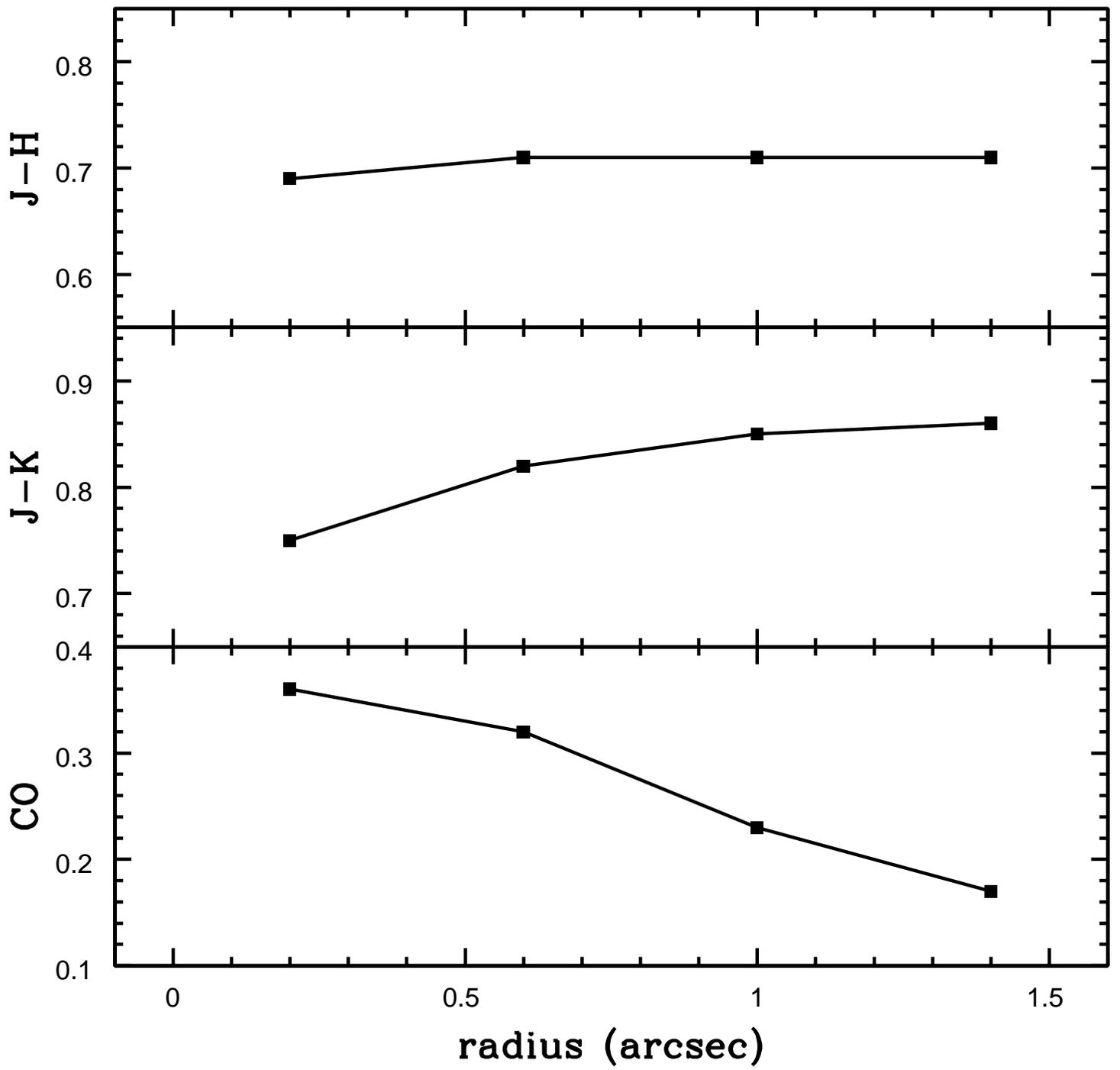
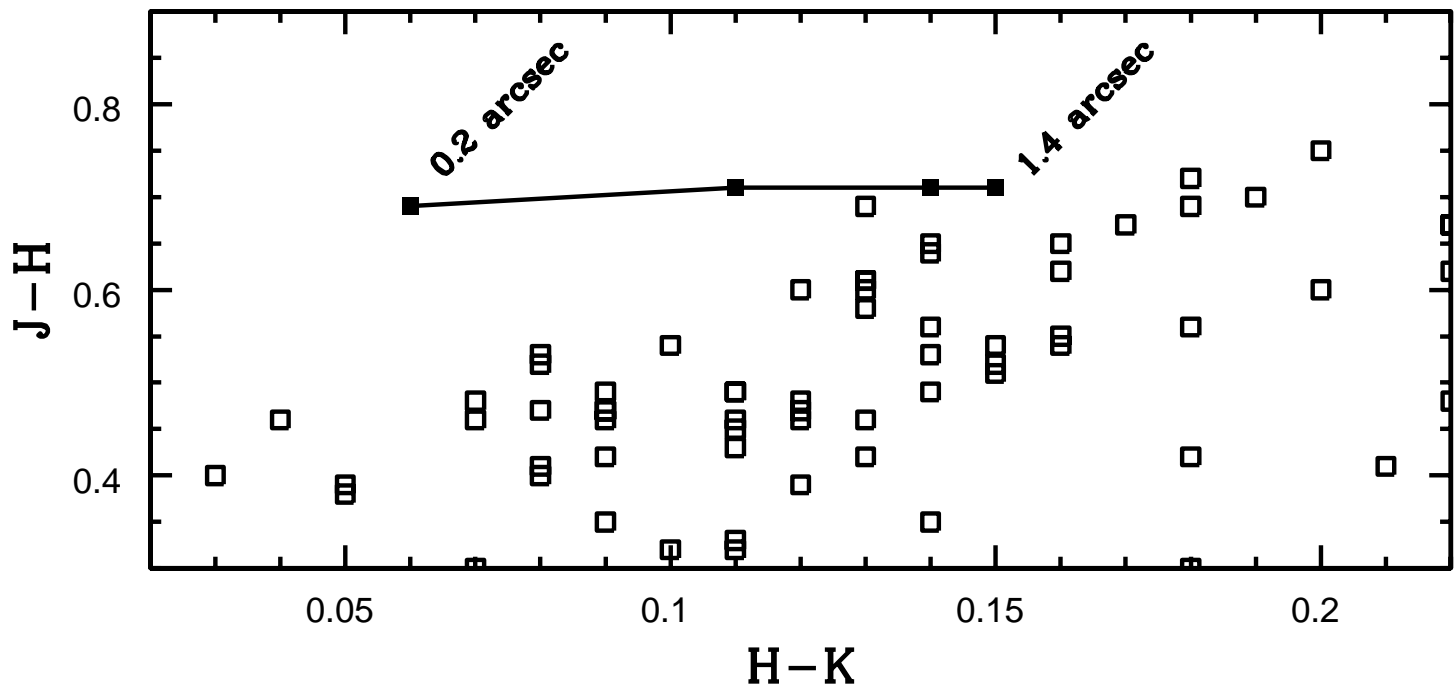
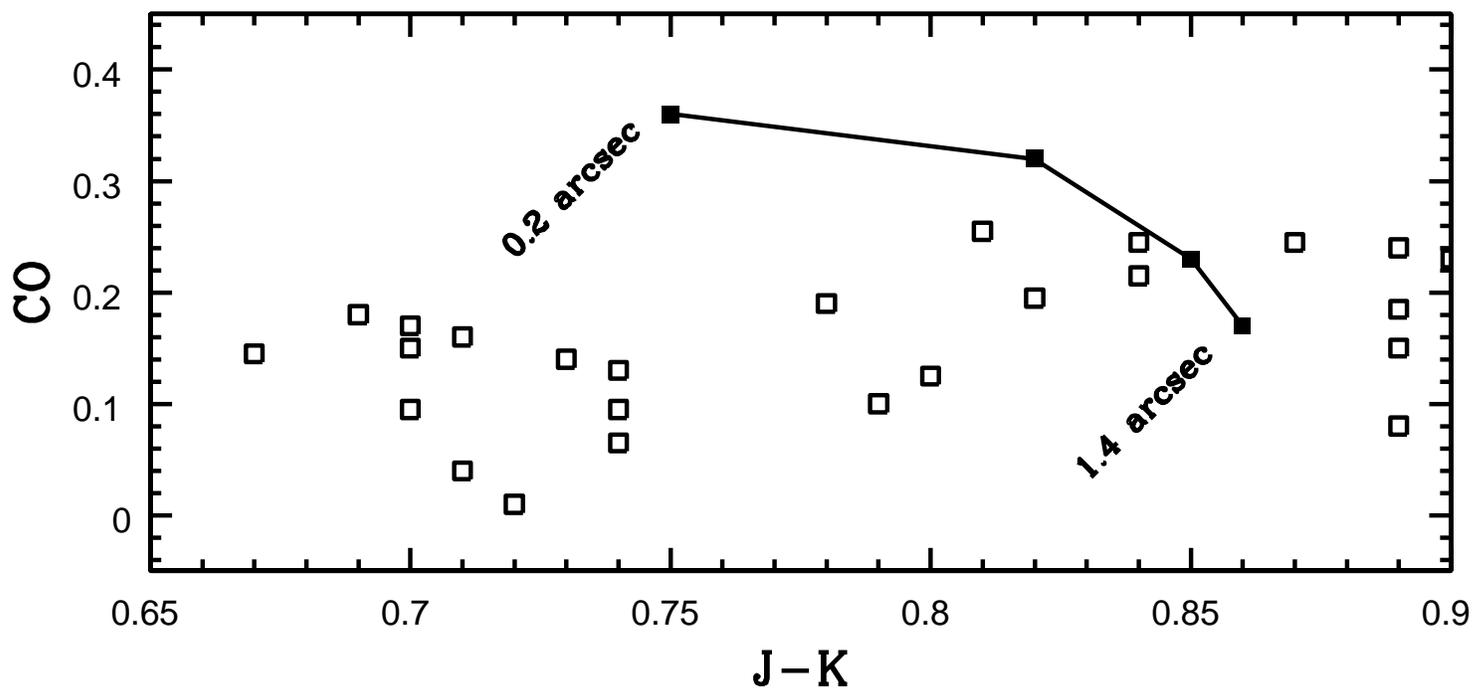
\end{document}